\documentclass[]{spie}

\usepackage{amsmath,amsfonts,amssymb}
\usepackage{graphicx}
\usepackage[colorlinks=true, allcolors=blue]{hyperref}

\title{Laser Metrology for Precision Alignment of Transmission Gratings in the REDSoX Soft X-ray Polarimeter}

\author[a]{Swati Ravi}
\author[a]{Alan Garner}
\author[a]{Jill Juneau}
\author[a]{F. Elio Angile}
\author[a]{Ralf K. Heilmann}
\author[a]{Herman L. Marshall}
\author[a]{Sarah N.T. Heine}
\affil[a]{MIT Kavli Institute for Astrophysics and Space Research, Massachusetts Institute of Technology, 77 Massachusetts Avenue, Cambridge, MA 02139, USA}

\authorinfo{Email: swatir@mit.edu}

\begin{document} 
\maketitle

\begin{abstract}
The Rocket Experiment Demonstration of a Soft X-ray Polarimeter (REDSoX) is a NASA sounding-rocket mission designed to perform the first astrophysical spectropolarimetry in the 0.2--0.4~keV energy band. The instrument uses critical-angle transmission (CAT) gratings to disperse incident X-rays onto laterally graded multilayer (LGML) mirrors, requiring 48 individual gratings to be co-aligned to within 6~arcminutes in yaw, pitch, and roll. To support the systematic assembly of the grating array, we adapted a scanning laser-reflection metrology technique in which normal-reflected, angled-reflected, and diffracted ultraviolet laser beams are measured using three position-sensitive detectors (PSDs). Changes in the measured beam positions are used to reconstruct the local yaw, pitch, and roll of each grating and provide real-time feedback during mechanical adjustment. We demonstrate the system using a prototype miniature grating structure containing two gratings. Following co-alignment, the assembly underwent a flight-level random-vibration test followed by a qualification-level sine sweep. Measurements obtained before and after testing showed that the relative grating orientations were retained to within 1~arcminute, less than 17\% of the REDSoX co-alignment tolerance. This work establishes a reproducible and scalable approach to the assembly and verification of large transmission-grating arrays for REDSoX and future X-ray spectroscopic instruments.
\end{abstract}

\keywords{X-ray spectroscopy, gratings, polarimetry, X-ray astronomy}

\section{INTRODUCTION}
\label{sec:intro} 
X-ray polarimetry provides information about the geometry, magnetic fields, and scattering environments of astrophysical sources that cannot be obtained from spectroscopy or timing alone. The Imaging X-ray Polarimetry Explorer \cite{2021AJ....162..208S, 2022JATIS...8b6002W} has opened the 2--8~keV band to routine polarimetric observations, but comparable measurements below 1~keV remain unavailable. The Rocket Experiment Demonstration of a Soft X-ray Polarimeter (REDSoX) is a NASA sounding-rocket mission designed to extend astrophysical spectropolarimetry into the 0.2--0.4~keV energy range \cite{2017SPIE10397E..0KM,2025SPIE13625E..10T, Heine2026REDSoX}. Its flight, planned for late 2027, will observe the blazar Markarian 421 while demonstrating technologies required for future orbital soft X-ray polarimeters. REDSoX also serves as a pathfinder for the Globe-Orbiting Soft X-ray Polarimeter (GOSoX), a NASA Pioneer mission based on the same instrument concept \cite{Marshall2026GOSoX}.

\begin{figure}[htbp]
    \centering
    \includegraphics[width=0.95\linewidth]{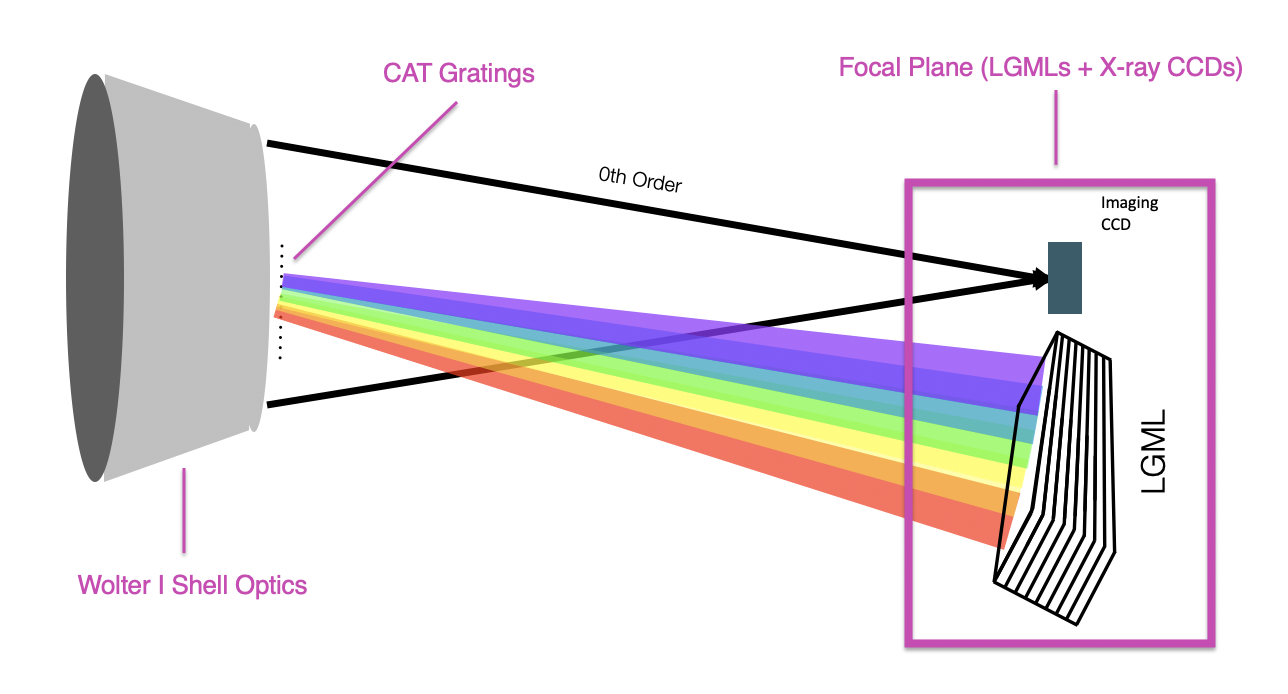}
    \caption{Schematic of the REDSoX optical system. Wolter-I shell optics focus incident X-rays through CAT gratings, which disperse the radiation onto LGMLs in the focal-plane assembly. The LGMLs reflect polarization-dependent components of the dispersed radiation toward the X-ray detectors (not shown).}
    \label{fig:redsox_instrument}
\end{figure}

REDSoX measures linear polarization using a sequence of Wolter-I focusing optics \cite{2025SPIE13626E..0PB}, critical-angle transmission (CAT) gratings \cite{2025SPIE13626E..0IH}, laterally graded multilayer (LGML) mirrors \cite{2015SPIE.9603E..19M}, and charge-coupled-device (CCD) detectors \cite{2024SPIE13103E..16H}, as shown schematically in Figure~\ref{fig:redsox_instrument}. The focusing optics direct incident X-rays through six ``petals'' (mounting arrays) of CAT gratings, which disperse the light by wavelength. The dispersed spectrum is incident on LGML mirrors mounted at approximately $45^\circ$, near Brewster's angle, where the reflectivity depends strongly on the orientation of the incident electric field. Three polarimetric channels oriented $120^\circ$ apart allow the linear polarization degree and angle to be reconstructed as functions of energy. The multilayer period of each LGML varies along its surface so that its local Bragg-response energy matches the wavelength dispersed to that position by the gratings \cite{2015SPIE.9603E..19M}. Consequently, the spectral and polarimetric response of the instrument depends on maintaining the prescribed correspondence between grating-dispersed wavelength and position on the LGML.

The REDSoX grating assembly contains 48 CAT gratings distributed among six petals \cite{Heine2026REDSoX}, described in greater detail in section \ref{subsec: grating module}. Small relative offsets of the gratings shift the location and orientation of the dispersed spectra, reducing their overlap with the corresponding LGML response and degrading the effective area of the instrument. The gratings must therefore be co-aligned to within 6~arcminutes in yaw, pitch, and roll. Achieving this tolerance across 48 separately mounted gratings requires a metrology system that can precisely measure all three angular degrees of freedom reproducibly while providing sufficiently rapid feedback to guide real-time adjustment during assembly.

Laser-reflection metrology offers a non-contact means of measuring grating orientation in air. Previous work developed a scanning laser-reflection tool for characterizing CAT-grating period variations and relative angular alignment \cite{2017SPIE10399E..15S}. In that implementation, the positions of normal-reflected, angled-reflected, and diffracted ultraviolet laser beams were measured using position-sensitive detectors (PSDs), allowing local grating yaw, pitch, and roll to be reconstructed. The method demonstrated arcminute-level angular accuracy and was subsequently applied to CAT-grating alignment and testing for soft X-ray instrumentation \cite{2021SPIE11444E..5ZG}.

In this work, we adapt this technique into a practical alignment system for the REDSoX grating-petal architecture. The system continuously converts measured beam positions into the relative yaw, pitch, and roll of mounted gratings, providing real-time feedback as their orientations are mechanically adjusted. We describe the REDSoX alignment requirements, the optical and mechanical configuration of the metrology system, and the geometric reconstruction used to determine the three alignment angles. We then demonstrate the procedure using a miniature flight-like structure containing two gratings and evaluate retention of their relative alignment following a flight-level random-vibration test and qualification-level sine sweep. 

\section{REDSoX grating assembly and alignment requirements}
\subsection{Grating module and petal architecture} \label{subsec: grating module}

\begin{figure}[htbp]
    \centering

    \begin{minipage}[t]{0.40\linewidth}
        \centering
        \includegraphics[width=\linewidth]{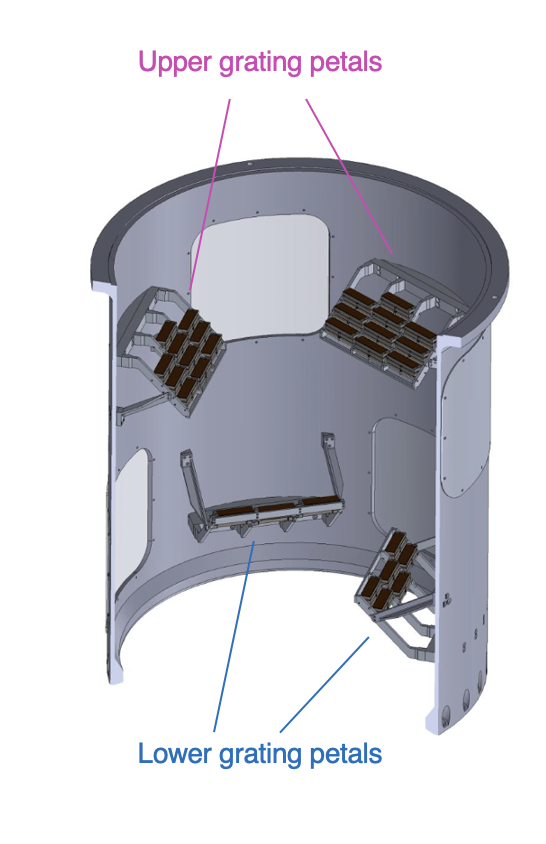}

        (a)
    \end{minipage}
    \hfill
    \begin{minipage}[t]{0.50\linewidth}
        \centering
        \includegraphics[width=\linewidth]{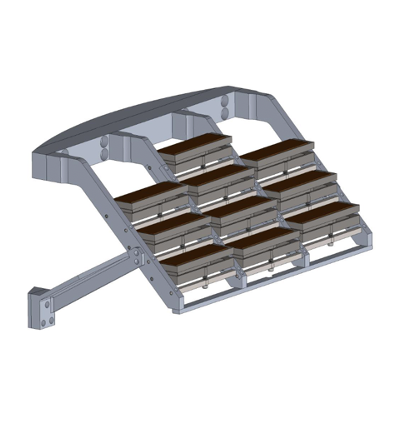}

        (b)
    \end{minipage}
    \vspace{0.5em}
    \caption{REDSoX grating-module architecture. (a) Cylindrical grating module, shown with four of its six grating petals installed. An upper petal and the lower petal rotated \(180^\circ\) from it form one of the instrument's three polarimetry channels; for example, the leftmost upper petal and rightmost lower petal shown here form one channel. (b) Upper
    grating petal containing ten individually mounted CAT gratings. The six petals contain a total of 48 gratings, with ten gratings  in each upper petal and six gratings in each lower petal.}
    \label{fig:grating_module}
\end{figure}

The REDSoX grating module contains 48 CAT gratings arranged among six mounting petals. Each of the instrument’s three polarimetric channels contains an upper and lower grating petal positioned in the converging beam from the Wolter-I optics. The gratings disperse incident X-rays toward a corresponding LGML in the focal-plane assembly. Depending on its location within the converging beam (upper or lower), each petal supports either ten or six gratings, respectively. Figure~\ref{fig:grating_module} shows the arrangement of the petals within the cylindrical grating module and the architecture of an individual upper petal.

Each CAT grating has a period of 200~nm and an active area of approximately $1\times3$~cm. The high-aspect-ratio silicon grating bars are supported by an integrated hierarchy of support structures and an outer silicon frame \cite{2025SPIE13626E..0IH}. Each grating is bonded to an individual titanium mount, which is installed in the larger petal structure. Adjustment screws accessible from the rear of the petal provide controlled changes in the grating orientation. The petal is attached to the cylindrical grating module through a replaceable mounting plate, allowing its final position to be matched to the measured multilayer gradient without requiring the aligned gratings to be removed from the petal \cite{Heine2026REDSoX}.

The modular architecture permits individual gratings to be characterized, mounted, and aligned before the completed petal is integrated into the payload. Such modularity is particularly valuable because each REDSoX grating is individually characterized and exhibits variation in efficiency \cite{Heine2026REDSoX}, allowing the measured performance of the available gratings to guide their final selection and placement within the petal assembly. This modular design, however, introduces a significant assembly challenge: the orientations of as many as ten separately mounted gratings must be made mutually consistent while preserving the prescribed orientation of the petal relative to the remainder of the optical system. A repeatable, non-contact measurement of each grating’s orientation is therefore required for assembly and integration of the REDSoX grating module.

\subsection{Grating orientation and co-alignment requirements}

\begin{figure}[htbp]
    \centering
    \includegraphics[width=0.75\linewidth]
    {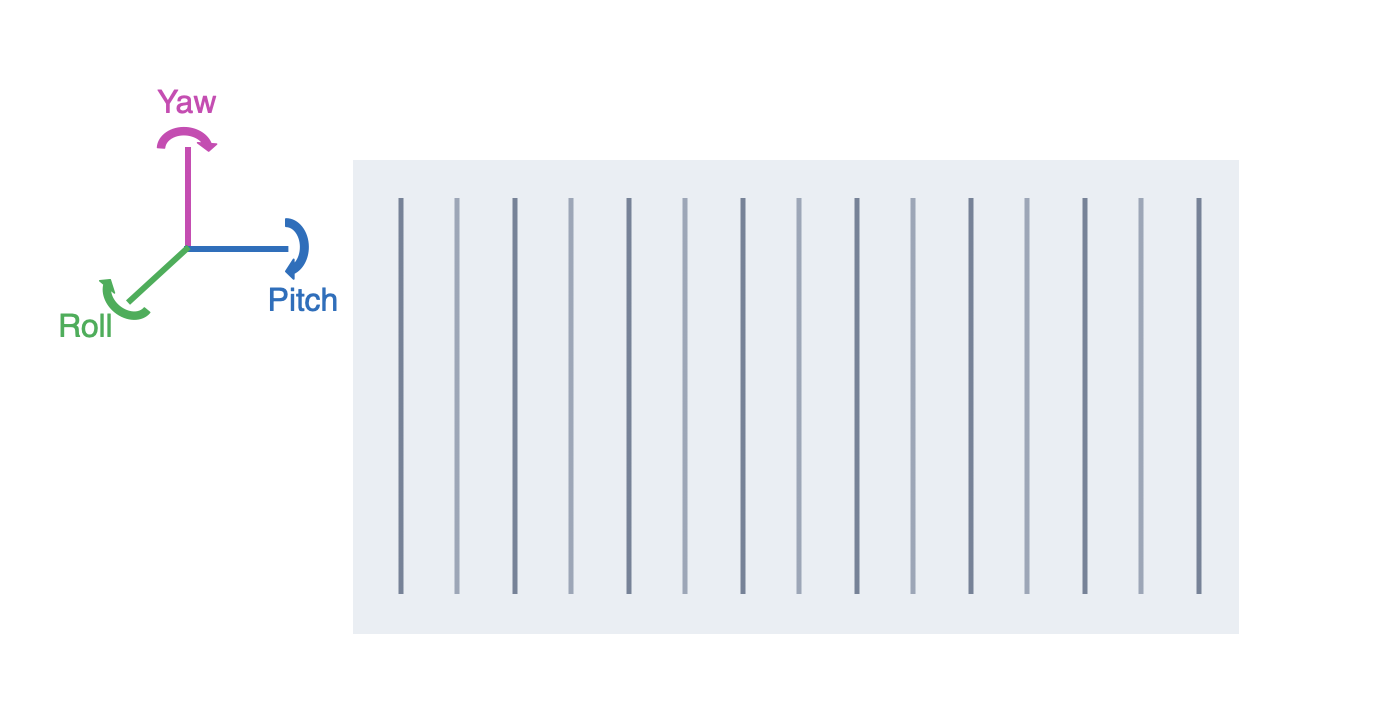}
    \caption{Definition of the local pitch, yaw, and roll angles used to describe CAT-grating orientation. The grating is represented face-on along its surface normal, with its long side along the \(x\)-axis and its short side along the \(y\)-axis. The lines parallel to the \(y\)-axis schematically represent the CAT-grating bars; their spacing and number are not shown to scale. Pitch is a rotation about the in-plane \(x\)-axis, yaw is a rotation about the orthogonal in-plane \(y\)-axis, and roll is a rotation about the surface normal pointing out of the page.}
    \label{fig:angle_definitions}
\end{figure}

The orientation of each grating is described by three local rotations: yaw, pitch, and roll. We define yaw and pitch as rotations of the grating surface normal about two orthogonal axes in the grating plane. Roll is a rotation about the surface normal and therefore describes the orientation of the grating bars within the grating plane. The sign conventions and corresponding grating coordinate system used throughout this work are illustrated in Figure~\ref{fig:angle_definitions}.

These angular degrees of freedom affect the REDSoX response in different ways. Because the CAT gratings operate in transmission, small pitch and yaw rotations do not directly produce an equivalent rotation of the diffracted beam in the laboratory frame, as would occur for a reflection grating. Instead, pitch and yaw primarily change the axial position of different regions of the grating surface and therefore the local grating-to-LGML throw distance. Photons diffracted from these regions consequently intercept the LGML at shifted positions. These rotations can also alter the local incidence, or blaze, angle and thereby redistribute photons among diffraction orders \cite{2021SPIE11444E..60G}. Roll rotates the grating bars within the grating plane and therefore rotates the dispersion direction, producing a transverse displacement of the spectral trace. Because the multilayer period of each LGML varies along its surface, photons of a given wavelength must arrive at the position where the local multilayer spacing satisfies the Bragg condition. Grating-orientation errors can disrupt this correspondence between wavelength (\(\lambda\)) and multilayer spacing (\(d\)), reducing the reflected intensity and therefore the effective area of the associated polarimetric channel.

The alignment budget requires the gratings within the array to be co-aligned to within 6~arcminutes in each of yaw, pitch, and roll. This relative co-alignment is more stringent than several of the module-level mechanical tolerances and is among the most demanding alignment tasks in the REDSoX payload. It is also distinct from the placement of the completed petal within the grating module: the metrology system described here measures the local angular orientation of individual gratings, whereas the position and orientation of the completed petals relative to the other instrument subsystems are established during payload integration.

\subsection{Two-grating mini-petal prototype structure and mounting procedure} \label{subsec: mini-petal}

\begin{figure}[htbp]
    \centering
    \begin{minipage}[t]{0.42\linewidth}
        \centering
        \includegraphics[width=\linewidth]{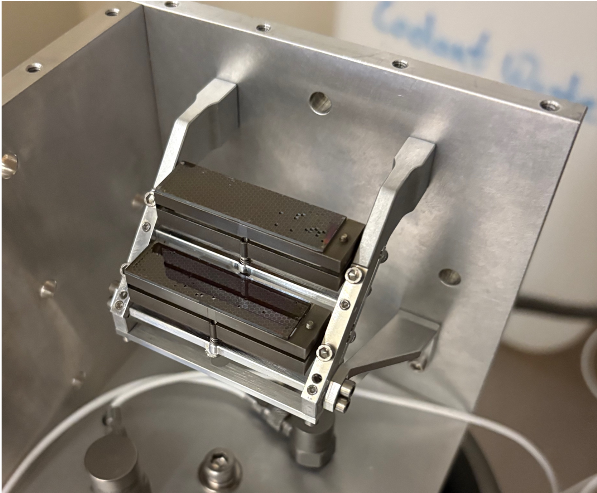}
        (a)
    \end{minipage}
    \hfill
    \begin{minipage}[t]{0.54\linewidth}
        \centering
        \includegraphics[width=\linewidth]{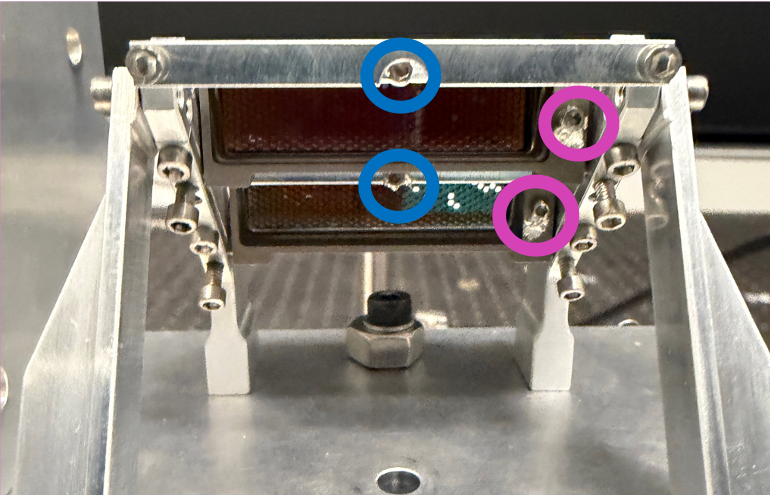}
        (b)
    \end{minipage}

    \vspace{0.5em}

    \begin{minipage}[t]{0.48\linewidth}
        \centering
        \includegraphics[width=\linewidth]{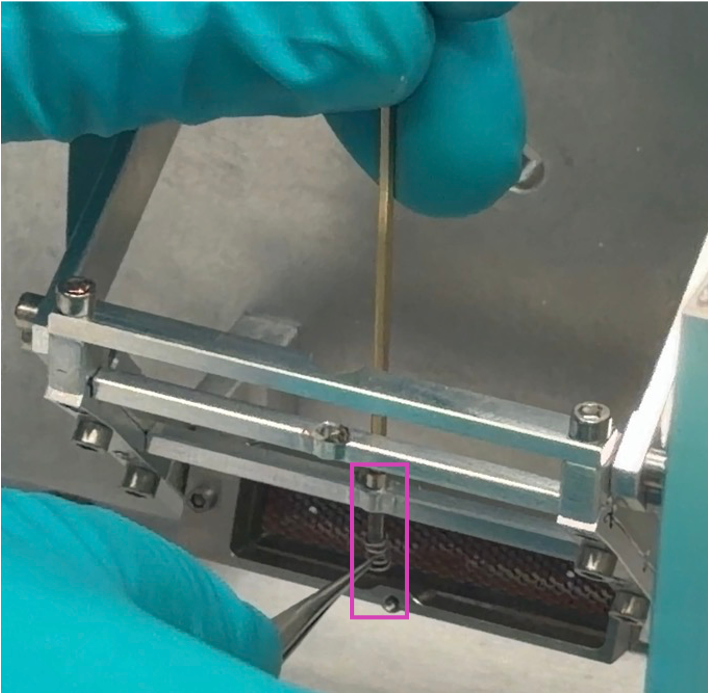}

        (c)
    \end{minipage}
    \hfill
    \begin{minipage}[t]{0.48\linewidth}
        \centering
        \includegraphics[width=\linewidth]{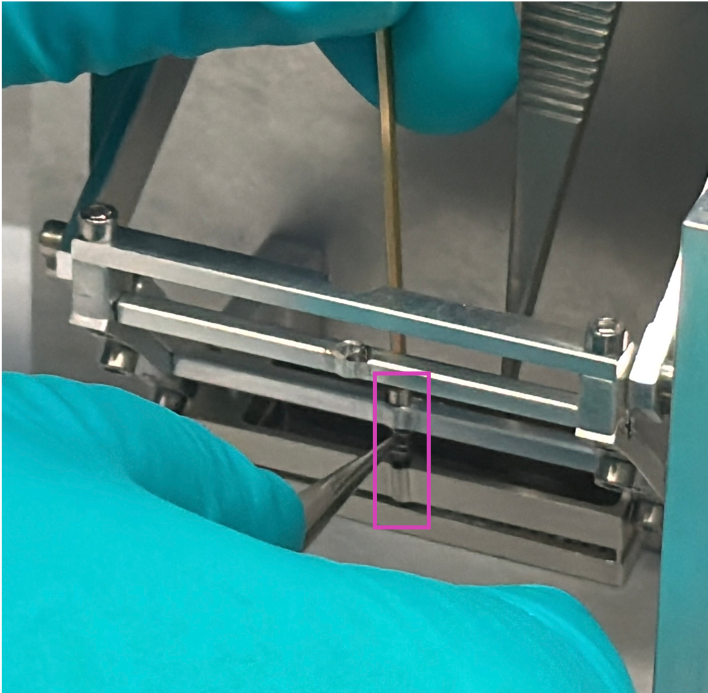}

        (d)
    \end{minipage}
    \vspace{0.5em}
    \caption{Two-grating mini-petal prototype and adjustment/mounting procedure used to demonstrate REDSoX grating co-alignment. (a) Front view of the mini-petal containing two mounted CAT gratings. (b) Rear view showing the pitch- and yaw-adjustment hardware circled in blue and magenta, respectively. (c) The first grating hanging temporarily from lateral bars incorporated into its titanium mount while the pitch-adjustment screw and uncompressed spring (boxed in magenta) are positioned. (d) One operator supports and raises the grating mount while the second operator uses tweezers to retain the spring and engage the pitch-adjustment screw (boxed in magenta). Installation of subsequent gratings is more constrained because the occupied neighboring position prevents the mount from hanging freely; the spring must instead be compressed with tweezers and inserted into the narrow gap between the titanium mount and mini-petal before the screw is installed.}
    \label{fig:minipetal}
\end{figure}

Before alignment of the full flight petals, the metrology and adjustment procedure was demonstrated using a miniature prototype structure (``mini-petal'') containing two CAT gratings. The prototype reproduced the relevant grating-mounting and adjustment features of a flight petal while leaving the gratings and adjustment screws accessible on the optical bench. One grating served as the angular reference, and the orientation of the second was adjusted relative to it. The rear of the structure provided separate mechanical controls for grating pitch and yaw, as shown in Figure~\ref{fig:minipetal}.

Installation of a grating mount in the mini-petal is a two-person procedure because the grating must be supported while the spring and pitch-adjustment screw are positioned. The first grating is comparatively straightforward to install: lateral bars incorporated into the titanium grating mount allow it to hang temporarily from the mini-petal. The pitch-adjustment screw is passed through the mini-petal, and a small spring is placed over the screw without initially applying tension. One operator then supports the grating mount in its final position while a second operator holds the spring in place and threads the pitch-adjustment screw into the titanium mount.

Installation becomes more constrained after the first grating occupies the mini-petal. The remaining clearance between the edge of the mini-petal and the adjacent titanium grating mount is insufficient for the next mount to hang freely from its lateral bars. The second grating must therefore be supported continuously while the spring and pitch-adjustment screw are inserted into the narrow available gap. The spring is temporarily compressed to provide sufficient clearance for the screw to engage the titanium mount. This procedure requires coordinated handling by two operators to maintain control of the grating while positioning the adjustment hardware.

The restricted access encountered in the two-grating mini-petal prototype illustrates an additional challenge for the assembly of flight petals containing six to ten gratings. The mounting and alignment sequence must be planned so that each additional grating can be installed without disturbing or obstructing access to those already in place. The non-contact nature of the laser-metrology system is advantageous in this configuration because grating alignment can be evaluated without introducing additional mechanical access around the mounted gratings.

The mini-petal was used to demonstrate the complete alignment procedure and to evaluate alignment retention under launch-like mechanical loading. These tests and their results are presented in Section~\ref{sec:env}.

\section{Laser Metrology System}

\subsection{Optical Configuration} \label{subsec: optical configuration}
\begin{figure}[htbp]
    \centering
    \includegraphics[width=0.95\linewidth]{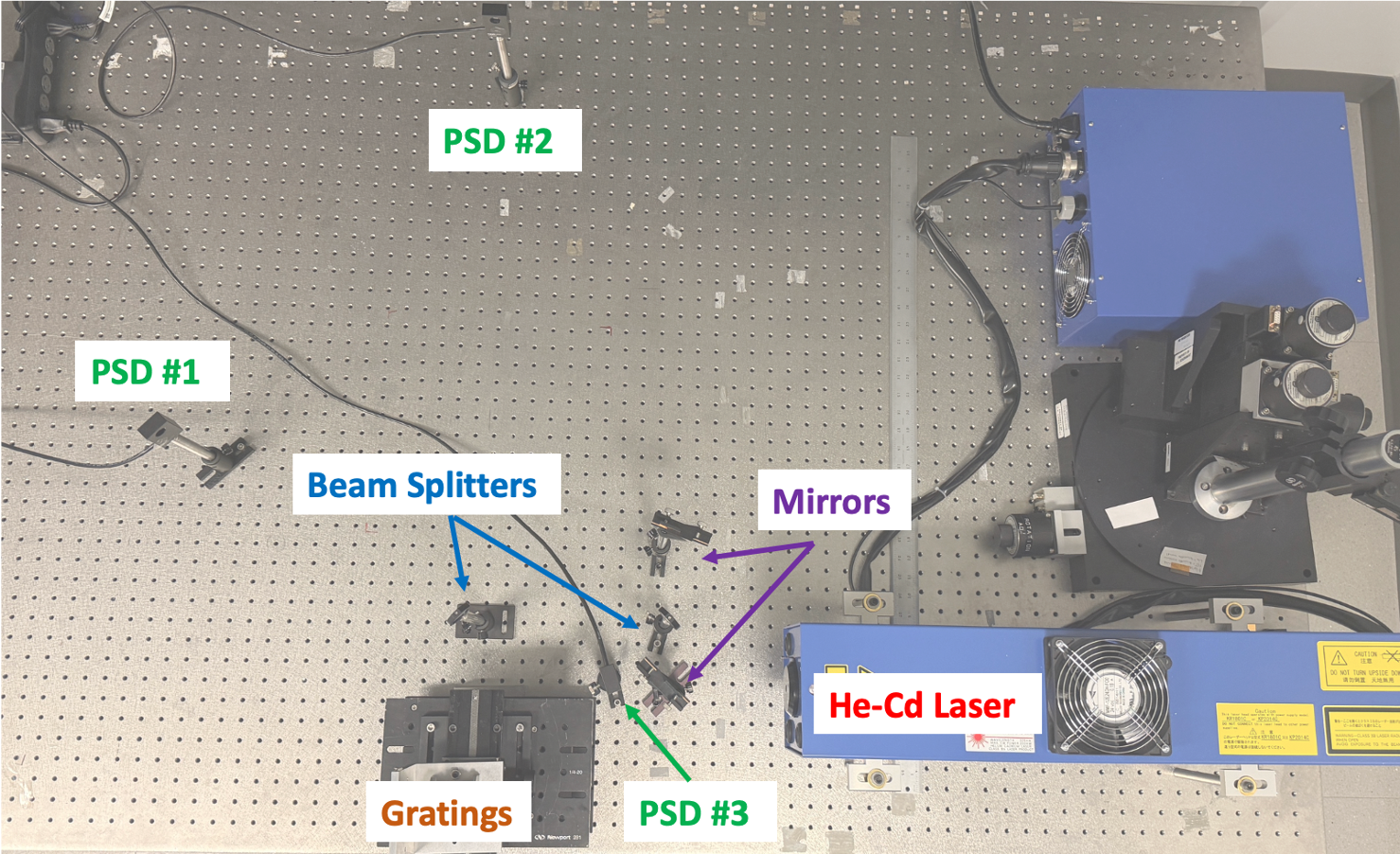}
    \caption{Optical-bench configuration of the REDSoX grating co-alignment laser-metrology system. A 325-nm helium-cadmium laser, beam splitters, and mirrors illuminate the mounted CAT grating along normal- and angled-incidence paths. Three position-sensitive detectors measure the normal-reflected, angled-reflected, and diffracted laser beams. These beam paths are illustrated in Figure~\ref{fig:beam_geometry}.}
    \label{fig:metrology_bench}
\end{figure}

The grating-alignment system adapts the scanning laser-reflection technique developed for characterizing CAT grating period variation \cite{2017SPIE10399E..15S} to the geometry and assembly requirements of the REDSoX grating petals. A 325-nm helium-cadmium laser provides a non-contact optical probe of the mounted grating surface and bar orientation. Beam splitters and mirrors direct the laser onto the grating along normal- and angled-incidence paths. The resulting normal-reflected, angled-reflected, and diffracted beams are directed onto three PSDs, as shown in Figures~\ref{fig:metrology_bench} and \ref{fig:beam_geometry}. Each PSD provides the two-dimensional centroid of the incident laser spot, allowing small changes in beam position to be measured continuously as the grating orientation is adjusted. During alignment, the mini-petal is mounted on a three-axis translation stage, allowing each grating to be positioned sequentially within the fixed incident beam without reconfiguring the laser-metrology optics or detectors.

The normal-reflected beam is primarily sensitive to rotations of the grating surface normal, while the diffracted beam is sensitive to both the surface orientation and the direction of the grating bars. These two beams provide the measurements used in the angular reconstruction presented here. The angled-reflected beam provides complementary monitoring of the reflected-beam geometry. The apparatus measures angular differences relative to a selected reference position rather than directly establishing an absolute orientation.

The optical components are mounted on a bench and the beam paths are arranged so that the relevant reflected and diffracted spots remain within the active areas of their respective PSDs throughout the range of mechanical adjustment. The distances from the illuminated grating to the normal-reflection and diffraction detectors are denoted by $R_{N}$ and $R_{D}$, respectively. The incident and diffracted beam angles are denoted by $\theta_i$ and $\theta_d$. These geometric quantities provide the conversion between measured linear displacements at the PSDs and angular changes at the grating.

\subsection{Reconstruction of grating orientation}

\begin{figure}[htbp]
    \centering
    \includegraphics[width=0.55\linewidth]{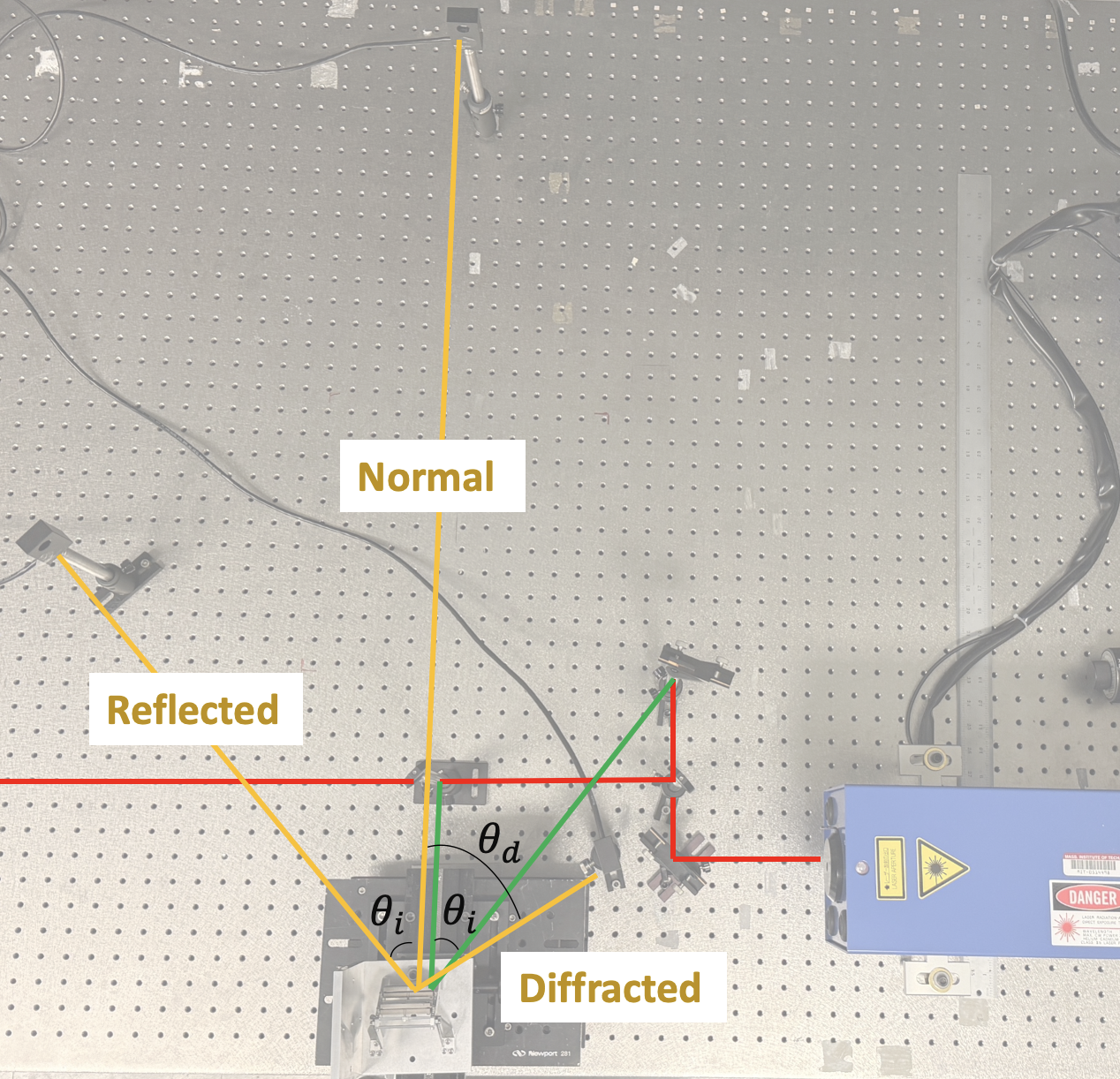}
    \caption{Geometry used to reconstruct CAT-grating orientation from the measured laser beams, overlaid on the optical-bench setup. The incident laser path is shown in two stages, first in red and then in green, while the normal-reflected, angled-reflected, and diffracted beams are shown in gold. Displacements \(dX_N\) and \(dY_N\) of the normal-reflected beam constrain yaw and pitch, respectively. The displacement \(dY_D\) of the diffracted beam, together with the measured pitch and the incident and diffracted angles \(\theta_i\) and \(\theta_d\), constrains roll. The corresponding grating-to-detector distances are \(R_N\) and \(R_D\). The normal-reflected and diffracted beam positions are used in the angular reconstruction presented here; the angled-reflected beam provides complementary monitoring of the reflected-beam geometry.}
    \label{fig:beam_geometry}
\end{figure}

The angular reconstruction follows the geometry developed for the CAT-grating scanning laser-reflection tool \cite{2017SPIE10399E..15S}. Figure~\ref{fig:beam_geometry} defines the beam-path lengths and incident and diffraction angles used below. For small angular changes, a rotation of a reflective surface produces a change in reflected-beam direction equal to twice the surface rotation. The axes defined in Figure~\ref{fig:angle_definitions} lie in the grating plane and are distinct from the detector coordinates \(X_N\) and \(Y_N\) used below. The displacement of the normal-reflected beam therefore provides direct measurements of grating yaw and pitch:

\begin{equation}
d\mathrm{Yaw} = \frac{dX_N}{2R_N},
\label{eq 1}
\end{equation}

\begin{equation}
d\mathrm{Pitch} = -\frac{dY_N}{2R_N},
\label{eq 2}
\end{equation}

\noindent where $dX_{N}$ and $dY_{N}$ are the changes in the horizontal and vertical positions, respectively, of the normal-reflected beam on its PSD. 

A change in the vertical position of the diffracted beam depends on both the surface pitch and the orientation of the grating bars. After accounting for the pitch contribution determined from the normal-reflected beam, the relative roll is given by

\begin{equation}
d\mathrm{Roll} =
-\frac{
dY_D + R_D(\cos\theta_d+\cos\theta_i)d\mathrm{Pitch}
}{
R_D(\sin\theta_d+\sin\theta_i)
},
\label{eq 3}
\end{equation}

where $dY_{D}$ is the measured vertical displacement of the diffracted beam. Equations~\ref{eq 1}--\ref{eq 3} are the linearized forms used for real-time alignment. They assume that the angular changes are small and that higher-order contributions from changes in surface height are negligible over the local measurement. These assumptions are appropriate for measuring the relative orientation of gratings that have already been installed near their nominal positions.

The reconstruction determines changes relative to an initial or reference grating orientation. For co-alignment, the beam positions measured from one grating are adopted as the reference, and the corresponding beam positions from each additional grating are converted into relative yaw, pitch, and roll offsets. The goal is therefore not to force the three PSD centroids to a predetermined absolute location, but to make the reconstructed orientation of each grating consistent with that of the selected reference grating to within the 6-arcminute requirement.

\subsection{Data Acquisition and Real-Time Feedback} \label{subsec:feedback}

The PSD signals are sampled at 10~Hz by custom acquisition software. Every individual measurement is recorded and saved, preserving the full temporal resolution of the PSD data. To provide a stable and readable display during alignment, the console is updated after each set of ten measurements, reporting the beam positions and reconstructed yaw, pitch, and roll from the tenth measurement acquired during the preceding one-second interval. 

The reconstructed angles are displayed while an operator adjusts the mounted grating. This real-time feedback allows the direction and approximate magnitude of each mechanical correction to be evaluated without removing the grating assembly or interrupting the alignment procedure. Retaining the individual 10-Hz measurements also allows short-timescale fluctuations and measurement stability to be examined after the alignment session rather than limiting the saved data to the one-second updates shown on the console.  

During alignment of the two-grating mini-petal prototype, the translation stage described in Section~\ref{subsec: optical configuration} was used to position each grating sequentially in the fixed incident beam. The normal-reflected, angled-reflected, and diffracted beam positions measured from the first grating were adopted as the reference. The stage was then translated manually to place the second grating in the same fixed incident laser beam without adjusting any of the laser-metrology optics or PSDs. The yaw, pitch, and roll of the second grating were calculated relative to the reference measurements. Its pitch- and yaw-adjustment screws were changed iteratively while the PSD readouts were monitored. This process continued until the measured angular differences between the two gratings satisfied the REDSoX co-alignment requirement. Translation-to-rotation coupling in the stage was not independently characterized during this demonstration and may contribute a systematic component to the measured relative angular offsets. A planned autocollimator measurement described in Section~\ref{subsec:absolute_alignment} will quantify this contribution before alignment of the flight petals.

Simultaneous monitoring of the three reconstructed angles is valuable because a mechanical adjustment intended to change one angle may also affect the others. Changes to the pitch- or yaw-adjustment screws produce corresponding displacements of the normal-reflected beam, while the diffracted-beam measurement reveals any accompanying change in roll. The system therefore allows the complete angular response of the grating to be monitored throughout the adjustment process, even though pitch and yaw are the actively controlled degrees of freedom.

The present implementation establishes grating orientations relative to a selected reference grating but does not independently define the absolute petal-alignment axis. The planned external reference axis is discussed in Section~\ref{subsec:absolute_alignment}.

\section{Mini-petal prototype alignment and environmental testing} \label{sec:env}

\subsection{Two-grating co-alignment demonstration}

The complete measurement and adjustment procedure was demonstrated using the two-grating mini-petal described in Section~\ref{subsec: mini-petal}. After both gratings were installed, the mini-petal was translated on a three-axis stage to place each grating sequentially in the fixed laser beam without adjusting the metrology optics or PSDs. The corresponding normal-reflected, angled-reflected, and diffracted beam positions were recorded. The reconstructed orientation of the first grating was adopted as the reference, and the yaw, pitch, and roll of the second grating were calculated relative to it.

\begin{figure}[htbp]
    \centering
    \includegraphics[width=0.55\linewidth]{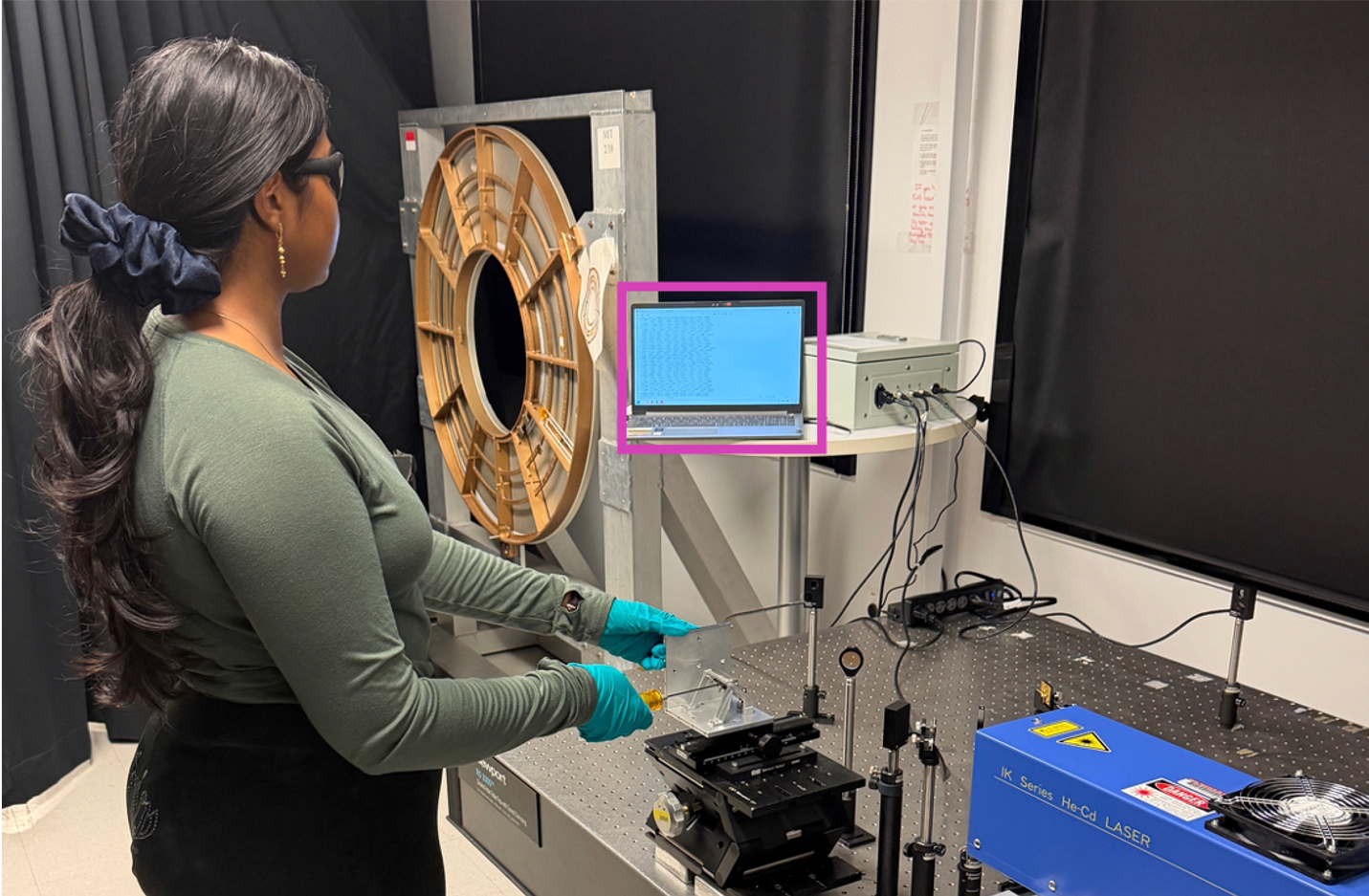}
    \caption{Real-time co-alignment of the two-grating mini-petal prototype. PSD measurements are acquired and saved at 10~Hz, and the corresponding pitch, yaw, and roll angles are calculated at the same cadence. The console (magenta box) displays every tenth measurement and its calculated angles, updating every second. The operator tightens or loosens the pitch- and yaw-adjustment screws as shown until the displayed angular offsets fall within the required co-alignment tolerance.}
    \label{fig:adjustment}
\end{figure}

The pitch- and yaw-adjustment screws of the second grating were then changed iteratively while the reconstructed angles were monitored through the console readout. The PSD measurements were acquired and saved at 10 Hz. Reconstructed angles calculated from the PSD readouts generally remained stable to within an arcminute when no adjustments were made to the mini-petal. Figure~\ref{fig:adjustment} shows the mini-petal on the optical bench during this alignment process along with the console display of the reconstructed angles.

The measured beam positions changed reproducibly in response to adjustment of the grating mount, allowing the effects of the pitch and yaw controls to be followed in real time. Simultaneous reconstruction of all three angular coordinates also allowed coupled changes to be identified during adjustment. The two gratings were brought into agreement within the 6-arcminute REDSoX co-alignment requirement in yaw, pitch, and roll, subject to the unquantified translation-to-rotation coupling discussed in Section~\ref{subsec:feedback}. This demonstration established that the laser-metrology system could be used not only to measure a completed assembly, but also to provide direct feedback during the mechanical alignment process.

\subsection{Vibration-test sequence and post-vibration alignment evaluation}

After co-alignment, the mini-petal underwent environmental testing to evaluate whether the relative grating orientations would be retained under launch-like mechanical loading. The prototype was subjected first to a flight-level random-vibration test and then to a qualification-level sine sweep following the applicable requirements of the NASA Sounding Rockets User Handbook \cite{Burth2023SoundingRockets}. The test exercised the prototype structure and its grating-mount interfaces under mechanical loads representative of, and in the case of the qualification-level sweep exceeding, those expected during flight.

The environmental test provided an initial evaluation of alignment retention in the mini-petal architecture. While the mini-petal prototype contained only two gratings rather than the six or ten that will populate a flight petal, it reproduced the relevant individual grating mounts, adjustment interfaces, and supporting petal structure. The gratings remained mounted throughout the vibration sequence and were returned to the laser-metrology bench without intentional readjustment.

The relative yaw, pitch, and roll of the two gratings were remeasured after the vibration sequence using the same laser-metrology procedure applied before testing. Comparison of the pre- and post-test measurements showed that all three relative alignment angles were retained to within 1~arcminute, with an alignment uncertainty of 0.3~arcminutes per angle determined from repeat measurements of a stationary grating. The maximum measured change corresponds to no more than 17\% of the 6-arcminute co-alignment tolerance. No mechanical damage or loss of adjustment capability was observed, and the mini-petal remained within the REDSoX co-alignment requirement following the test.

\section{Discussion and future development}

\subsection{Performance relative to REDSoX requirements}

The two-grating demonstration establishes that the laser-metrology system can provide sufficiently precise and responsive feedback to support CAT-grating alignment in the REDSoX petal architecture. Because the measurements are non-contact and can be performed while the gratings remain installed, their orientations can be monitored without introducing additional mechanical access around an increasingly populated petal.

The maximum measured change following vibration testing was no greater than 1~arcminute, corresponding to no more than 17\% of the REDSoX co-alignment tolerance. This margin provides initial validation of the grating-mount and adjustment interfaces and demonstrates that the same optical system can support both assembly and post-environmental-test verification. The result should nevertheless be interpreted as a two-grating proof-of-concept; validation of a fully populated petal remains necessary.

The method also complements the modular construction of the grating assembly. Each CAT grating is characterized individually before integration, allowing measured variations in diffraction efficiency to guide its selection and placement within the six petals. This information is used to distribute the gratings so that the three polarimetric channels have comparable effective areas. The selected gratings can then be aligned in air without requiring X-ray illumination or vacuum operation. Separating efficiency characterization from angular alignment therefore permits the grating distribution to be optimized while maintaining a common assembly and verification procedure.

\subsection{Scaling from the prototype to flight petals}

The two-grating mini-petal prototype reproduces the principal mounting and adjustment interfaces of the flight design, but a populated REDSoX petal will contain either six or ten gratings depending on its position within the optical system. Scaling to the flight hardware introduces practical considerations that are not fully represented by the prototype. In particular, the clearance available for installing the spring and pitch-adjustment screw decreases as neighboring grating positions are occupied. The assembly order must be chosen so that each new grating can be installed while maintaining access to the adjustment hardware and avoiding contact with gratings that have already been aligned.

The alignment sequence must similarly prevent errors from accumulating across the petal. Rather than aligning each grating only to its nearest neighbor, measurements should remain tied to a common reference so that a series of individually small offsets does not produce a larger difference between the first and last gratings.

A fully populated engineering or flight petal will provide the next opportunity to evaluate the complete workflow, including mounting order, optical access, adjustment time, reproducibility, and alignment retention across all grating positions. Environmental testing of a populated petal will also be required to demonstrate that the behavior of the two-grating prototype remains representative when the full complement of mounts and gratings is installed.

\subsection{Absolute alignment and planned upgrades} \label{subsec:absolute_alignment}

The present system determines the orientation of each grating relative to a selected reference grating. This is sufficient to establish mutual co-alignment, but it does not independently define the absolute axis to which the first grating and completed petal should be aligned. A planned upgrade will incorporate a reference mirror whose surface normal defines a fixed external alignment axis. The gratings can then be aligned absolutely to the reference-mirror axis rather than simply co-aligned to each other. This will connect the local grating-alignment measurements to the absolute orientation required for integration of the completed petal into the REDSoX payload. Furthermore, before alignment of a fully populated petal, translation-stage-induced angular deviations will be characterized using an autocollimator and a reference mirror rigidly mounted to the stage, allowing any solid-body rotation of the petal during translation between grating positions to be quantified and verified to remain within the allowable alignment error.

Additionally, the acquisition software records and saves individual PSD measurements at 10~Hz while displaying every tenth instantaneous angular estimate, producing a 1-Hz console update. A planned software upgrade will preserve the 10-Hz acquisition and recording rate but replace each displayed instantaneous estimate with the average of the ten measurements acquired during the preceding second. Further software development will also focus on streamlining the transition between grating positions, maintaining a common reference throughout the alignment of a populated petal, and producing a complete record of the measured orientation of every grating. 

Future environmental tests will retain the individual pre- and post-test yaw, pitch, and roll measurements in addition to the overall alignment-verification result. Repeated measurements before and after each vibration axis will allow true mechanical shifts to be separated from measurement scatter and will provide a more complete quantitative assessment of alignment retention. Together, the reference mirror, expanded software workflow, and full-petal testing will advance the present proof-of-concept system into the alignment and verification process for the REDSoX flight grating assembly. The same approach is directly relevant to GOSoX and other future X-ray instruments that require co-alignment of many individually mounted gratings. As these arrays increase in size, scalable alignment methods that operate in air, provide immediate feedback, and preserve a traceable measurement history will become increasingly important.

The two-grating mini-petal demonstration establishes that the laser-metrology system can guide CAT-grating co-alignment in flight-like REDSoX hardware and verify alignment retention following environmental testing. Extension to fully populated petals, more rigorous verification and environmental testing, and the addition of an absolute reference axis will provide the basis for assembly and co-alignment of the complete 48-grating REDSoX array. A planned end-to-end X-ray test of REDSoX at NASA Marshall Space Flight Center prior to launch will provide X-ray verification of the completed alignment by confirming that the dispersed spectra reach the expected locations in the focal plane and properly match the LGML response.

\bibliography{report} 
\bibliographystyle{spiebib} 

\end{document}